\documentclass[prb,twocolumn,preprintnumbers,amsmath,amssymb]{revtex4}

\usepackage{epsfig,amsfonts}

\def\be{\begin{equation}}
\def\ee{\end{equation}}

\def\bea{\begin{eqnarray}}
\def\eea{\end{eqnarray}}

\def\e{\epsilon}

\def\a{\alpha}
\def\b{\beta}

\def\ben{\begin{enumerate}}
\def\een{\end{enumerate}}

\newcommand{\ket}[1]{\left| #1 \right>}

\begin{document}

\title{Gaudin models solver based on the Bethe ansatz/ordinary differential equations correspondence}
\author{Alexandre Faribault${}^1$, Omar El Araby${}^2$, Christoph Str\"{a}ter${}^1$, Vladimir Gritsev${}^2$ }
\affiliation{$^1$Physics Department, ASC and CeNS, Ludwig-Maximilians-Universit\"{a}t, 80333 M\"{u}nchen, Germany}
\affiliation{$^2$Physics Department, University of Fribourg, Chemin du Mus\'{e}e 3, 1700 Fribourg, Switzerland}

\date{\today}

\begin{abstract}

We present a numerical approach which allows the solving of Bethe equations whose solutions define the eigenstates of Gaudin models. By focusing on a new set of variables, the canceling divergences which occur for certain values of the coupling strength no longer appear explicitly. The problem is thus reduced to a set of quadratic algebraic equations.  The required inverse transformation can then be realized using only linear operations and a standard polynomial root finding algorithm. The method is applied to Richardson's fermionic pairing model, the central spin model and generalized Dicke model.

\end{abstract}

\maketitle

The correspondence between  Bethe Ansatz / Ordinary Differential Equation (ODE) has been found some time ago\cite{doreyfirst} on the basis of similarity between T-Q system of the Bethe Ansatz and functional relations including spectral determinants on the ODE side. For extensive review we refer to \cite{doreyreview}. The correspondence is related to the Langlands correspondence where the ODE part has to do with so-called Miura opers\cite{feigin}. It has many interesting links with conformal field theory, quasi-exact solvability, supersymmetry and PT-symmetric quantum mechanics. It was also used it in the context of physics of cold atoms and quantum impurity problem \cite{gritsev}. Here we develop further this remarkable correspondence to facilitate the solving of otherwise difficult-to-solve Gaudin systems and implement it to several physically interesting situations. 

Indeed, numerous integrable models derived from a generalized Gaudin algebra have been used to describe the properties of physical systems. The Richardson fermionic pairing Hamiltonian\cite{rs-62} has explained most properties of superconducting nanograins \cite{exp,dr-01}, numerous studies of the decoherence of a single electron spin trapped in a quantum dot rely on the central spin model \cite{centralspin} while the inhomogeneous Dicke models \cite{Dicke} have been used to study light-matter interaction in many cavity-based systems.

The quantum integrability of such systems brings major simplifications to the structure of their eigenstates. In a system containing $N$ degrees of freedom, they can be fully described by a number $M$ of complex parameters we call rapidities. $M$ is here a number of the same order as $N$ despite the exponentially large Hilbert space. The possible values of this set of parameters can be obtained by finding solutions to a set of $M$ coupled non-linear algebraic equation: the Bethe equations. However, analytically finding solutions to the Bethe equations remains impossible except in very specific cases. The problem is still numerically approachable but solving non-linear equations remains a challenge that only iterative methods can tackle.

While previous efforts in designing algorithms have shown promising results [\onlinecite{rsd-02,r-66,s-07,rnd-57,ded-06}], methods which are sufficiently fast and stable for systematically finding a large number of eigenstates remained elusive. As was shown by one of the authors, computing efficiently only a small fraction of the complete Hilbert space can be sufficient to access static \cite{fcc-08}, dynamical \cite{fcc-10} and even non-equilibrium dynamical properties of these systems \cite{fcc-09,fcc-09p} making such a method highly desirable.

This paper presents an algorithm which finds solutions to the system of equations with unprecedented speed and stability. The first section discusses general considerations related to the quantum systems discussed here. Section \ref{solving} describes in detail the method used to solve for a set of intermediate variables and we show in section \ref{inverting}, how one can recover the rapidities from these variables. Explicit numerical applications to the Richardson model, the generalized Dicke model and the central spin model are presented in section \ref{applications} and we conclude in the remaining section.

\section{The models}
\label{general}

Let us first introduce the generalized Gaudin algebra defined by the operators $ \mathrm{S}^x(\lambda_i), \mathrm{S}^y(\lambda_i),\mathrm{S}^z(\lambda_i)$ with $\lambda_i$ any complex number and the commutation rules they obey \cite{Gaudin,Ortiz}:

\bea
\left[\mathrm{S}^x(\lambda_i),\mathrm{S}^y(\lambda_j)\right] &=& i(Y(\lambda_i,\lambda_j) \mathrm{S}^z(\lambda_i)- X(\lambda_i,\lambda_j)\mathrm{S}^z(\lambda_j)),
\nonumber\\
\left[\mathrm{S}^y(\lambda_i),\mathrm{S}^z(\lambda_j)\right]  &=&i(Z(\lambda_i,\lambda_j) \mathrm{S}^x(\lambda_i)- Y(\lambda_i,\lambda_j)\mathrm{S}^x(\lambda_j)) ,
\nonumber\\
\left[\mathrm{S}^z(\lambda_i),\mathrm{S}^x(\lambda_j)\right] &=&i(X(\lambda_i,\lambda_j) \mathrm{S}^y(\lambda_i)- Z(\lambda_i,\lambda_j)\mathrm{S}^y(\lambda_j)) ,
\nonumber\\
\left[\mathrm{S}^\kappa(\lambda_i),\mathrm{S}^\kappa(\lambda_j)\right] &=& 0 \ \ \ \kappa=x,y,z.
\eea

In this paper, we deal with the rational family of Gaudin models for which

\bea
X(\lambda_i,\lambda_j)=Y(\lambda_i,\lambda_j) = Z(\lambda_i,\lambda_j) &=&  \frac{g}{\lambda_i-\lambda_j}.
\eea

The generalized Dicke model we also consider is however obtained from a trigonometric Gaudin model \cite{dicke_derive} defined by

\bea
X(\lambda_i,\lambda_j)&=&Y(\lambda_i,\lambda_j) = \frac{g}{\sin(\lambda_i-\lambda_j)}
\nonumber\\
Z(\lambda_i,\lambda_j) &=& g \cot(\lambda_i-\lambda_j).
\eea

The model is then derived by taking the large spin limit of a Holstein-Primakoff transformation for one of the degrees of freedom. The resulting limit shares more resemblance to the rational family of Gaudin models and simply constitutes its extension to include a bosonic degree of freedom. We will therefore indiscriminately use the term rational to also include models derived from this particular limit.

For any given realization of the Gaudin algebra, one can define a set of $N$ commuting operators $R_i$ allowing one to build exactly-solvable Hamiltonians

\bea
H = \sum_{i=1}^N \eta_i R_i,
\eea

\noindent for which the $R_i$ are evidently constants of motion. Eigenstates of these models are, for a given number of excitations $M$, all defined by the construction

\bea
\left|\left\{\lambda_1 ... \lambda_M\right\} \right>\propto \prod_i \mathrm{S}^+(\lambda_i) \left| 0 \right>.
\label{bethestate}
\eea

Here $\mathrm{S}^+(u) = \mathrm{S}^x(u)+ i  \mathrm{S}^y(u)$ is a family of generalized creation operators parametrized by the complex parameter $u$. Its explicit expression in terms of the fundamental operators defining a particular realization will be model dependent. The pseudovacuum $\left|0 \right>$ is defined as the lowest weight vector, i.e. $\mathrm{S}^-(u) \left|0 \right> = 0, $ and can also differ in distinct realizations of the Gaudin algebra.

States of the form (\ref{bethestate}) become eigenstates of the system provided the $M$ rapidities $\lambda_i$ are solution of a set of coupled non-linear algebraic equation: the Bethe equations. For rational models, these equations can be written, in general, as

\bea
F(\lambda_i)= \sum_{j\ne i} \frac{1}{\lambda_i-\lambda_j},
\label{eq:RGeq}
\eea

\noindent with $S^z(\lambda_i)\left| 0 \right> = F(\lambda_i) \left| 0 \right>$ defining the lowest weight function $F$.

One of the major difficulties in solving these equations numerically is the divergences which occur whenever two rapidities coincide. While they are cancelled by similarly diverging terms on the left hand side of the equation, it still has an important impact on numerical stability and computational speed. To circumvent these potential pitfalls, we introduce the function

\bea
\Lambda(z) \equiv \sum_{k=1}^M \frac{1}{z-\lambda_k} =  \frac{P'(z)}{P(z)},
\label{sdef}
\eea

\noindent where

\bea
P(z) = \prod_{k=1}^M (z-\lambda_k)
\eea

\noindent is the polynomial of degree $M$ whose $M$ roots correspond to the values of $\lambda_k$.

Since $\Lambda(z)$ obeys the following Riccati-type differential equation \cite{BA/ODE,BA/ODE2}

\bea
\frac{\partial \Lambda(z)}{\partial z} + \Lambda^2(z) &=& -\sum_{\alpha} \frac{1}{(z-\lambda_\a)^2} + \sum_{\alpha,\beta} \frac{1}{(z-\lambda_\a)(z-\lambda_\b)}
\nonumber\\ &=&
\sum_{\a\ne\b} \frac{2}{(z-\lambda_\a)(\lambda_\a-\lambda_\b)},
\eea

\noindent it is easy to show that when the set $\{\lambda_i\}$ is a solution of the Bethe equations we have

\bea
\Lambda'(z)+ \Lambda^2(z) -
\sum_{\a} \frac{2 F(\lambda_\a)}{(z-\lambda_\a)} &=& 0.
\label{eq1}
\eea

One can derive this last equation with respect to $z$ any number of times to write additional equalities:
\bea
&&\Lambda''(z)+ 2\Lambda(z)\Lambda'(z) +
\sum_{\a} \frac{2 F(\lambda_\a)}{(z-\lambda_\a)^2} = 0,
\nonumber\\
&&
\Lambda'''(z)+ 2\Lambda(z)\Lambda''(z)+2\Lambda'(z)^2-
\sum_{\a} \frac{4 F(\lambda_\a)}{(z-\lambda_\a)^3} = 0,
\nonumber\\
&&
\dots
\label{othereqs}
\eea

From this point on, we will assume a particular form for the $F$ function which, while not general, encompasses a wide variety \cite{Ortiz} of physically relevant realizations of the Gaudin algebra. We restrict ourselves to the following

\bea
F(\lambda_\a) = - \sum_{i=1}^N \frac{A_i}{(\epsilon_i-\lambda_\a)} + \frac{B}{2g} \lambda_\a +\frac{C}{2g}.
\label{eq:ffunction}
\eea

The exact physical nature of the parameters $g$ and $\e_i$ is highly model dependent, but in the cases treated in this paper it will be made explicit in section \ref{applications} when discussing their specifics. In (pseudo-)spin models, $A_i = \left|s_i\right|\Omega_i$ with $ \left|s_i\right|$ the norm of the local spin degree of freedom, while $\Omega_i$ is an integer relatable to the degeneracy, i.e. the number of elements of the set $\{\epsilon_j\}$ equal to $\epsilon_i$ . Consequently, every $A_i$ can then take any integer or half-integer value.

\section{Solving the system}
\label{solving}

From the previously found set of differential equations, we can
write a new set of algebraic equations by simply taking the limits
$z \to \e_j$ of Eqs. (\ref{eq1}) and (\ref{othereqs}). Using the
previous form of $F(\lambda_\a)$ (Eq. (\ref{eq:ffunction}))  we find

\bea
&& (1-2A_j)\Lambda'(\e_j)+ \Lambda^2(\e_j) + \frac{B}{g}M -  \frac{B\e_j + C}{g}\Lambda(\e_j)\nonumber\\&&
 \ \ \
+ \sum_{i \ne j }2A_i \frac{\Lambda(\e_j)-\Lambda(\e_i)}{\e_i-\e_j}  =0,
\label{eq1}
\eea
\bea
&&(1-A_j)\Lambda''(\e_j)+ 2\Lambda(\e_j)\Lambda'(\e_j)  -
\frac{B}{g} \Lambda(\e_j)\nonumber\\&&
 \ \ \  -  \frac{B\e_j + C}{g}\Lambda'(\e_j)
+\sum_{i \ne j }
2A_i\frac{\Lambda(\e_j)-\Lambda(\e_i) }{(\e_i-\e_j)^2 }
\nonumber\\&&
 \ \ \ +\Lambda'(\e_j)\sum_{i \ne j } \frac{2A_i }{\e_i-\e_j} = 0,\nonumber\\
\label{eq2}
\eea
\bea
&&
\left(1-\frac{2}{3}A_j\right)\Lambda'''(\e_j)+ 2\Lambda(\e_j)\Lambda''(\e_j)+2\Lambda'(\e_j)^2 \nonumber\\&&
 \ \ \ - 2 \frac{B}{g}\Lambda'(\e_j) -  \frac{B\e_j + C}{g}\Lambda''(\e_j) \nonumber\\&&
 \ \ \
 +\sum_{i \ne j }
4A_i\frac{\Lambda(\e_j)-\Lambda(\e_i) }{(\e_i-\e_j)^3 }
+\sum_{i \ne j } \frac{4A_i \Lambda'(\e_j)}{(\e_i-\e_j)^2} \nonumber\\&& \ \ \ +\sum_{i \ne j } \frac{2A_i \Lambda''(\e_j)}{\e_i-\e_j}  = 0,
\label{eq3}
\nonumber\\
\\
&&\dots
\nonumber
\eea

One can immediately see that for any non-degenerate spin 1/2 degree of freedom ($A_j= \frac{1}{2}$), the first differential equation reduces to a quadratic algebraic one which depends only on the set  of variables $\left\{\Lambda(\e_j)\right\}$. This is due to the canceling of the first term ($1-2A_j = 0$).

For $A_j = 1$, which can occur either due to a doubly degenerate spin 1/2 or a single spin 1, the two first equations form a quadratic system of equations depending on $\left\{\Lambda(\e_j)\right\}$ but also on the first derivative $\Lambda'(\e_j)$ evaluated at $\e_j$. Larger values of $A_j$ require additional equations, but it is always possible to write a closed coupled system of quadratic equations. It would depend on $\Lambda(\e_j)$ for variables with $A_j=\frac{1}{2}$, on both $\Lambda(\e_j),\Lambda'(\e_j)$ for variables with $A_j=1$, $\Lambda(\e_j),\Lambda'(\e_j),\Lambda''(\e_j)$ for variables with $A_j=\frac{3}{2}$, etc. The resulting system of equations is built by using the $n$ first equations above for any variables with $A_j=\frac{n}{2}$ and its solutions would give a one to one correspondence with the solutions of the Bethe equations and therefore the eigenstates of the system.

There is one caveat that readers should be aware of. While a spin $S > \frac{1}{2}$ or a set of $2S$ degenerate spin-$\frac{1}{2}$ lead to the same set of Bethe equations, only in the former case would the solutions to the Bethe equations give us the full Hilbert space. This is particularly simple to understand for a system containing only one spin $S=1$ or two degenerate spin-$\frac{1}{2}$. In the first case, the Hilbert space dimension is 3 while in the second it is 4 and the resulting Bethe equations have only 3 distinct solutions. In the degenerate case, the Bethe equations would only give us highest weight states (one can think of it as the $J=1$ triplet for two spins-$\frac{1}{2}$) and the remainder of the Hilbert space would need to be reconstructed by building the appropriate set of orthogonal states.

While in principle the system of quadratic equations can be solved for any spin or degeneracies, for the remainder of this paper will focus on the simplest case of non-degenerate spin 1/2 systems ($A_j=\frac{1}{2} \  \ \forall \ j$). In this case, the closed set of algebraic equations is given by the $N$ following equations:

\bea
 \Lambda^2(\e_j)
&=&
 \sum_{i\ne j}^N \frac{\Lambda(\e_j)-\Lambda(\epsilon_i)}{\e_j-\epsilon_i} -M \frac{B}{g}
+\left( \frac{B}{g}\e_j+ \frac{C}{g}\right)\Lambda(\e_j).\nonumber\\
 \label{sequ}
\eea

In the subspace $M<N$ this system of equations is larger than the original one (Eq. (\ref{eq:RGeq})). However, being quadratic and not having canceling divergences, makes it a much easier problem to tackle numerically.

We also note in passing that the values of $\Lambda(\epsilon_{j})$ determine the eigenvalues $r_j$ of the commuting Hamiltonians $R_{j}$. The eigenvalues of the transfer matrix, $\tau(\lambda)=\frac{1}{2}\mbox{Tr}({\bf \mathrm{S}}^{2})$ are given in terms of $F(\lambda)$ (see Eq.(\ref{eq:ffunction})) as follows:

\bea
\tau(\lambda)=F^{2}(\lambda)+F'(\lambda)+2\sum_{i}(F(\lambda)-F(\lambda_{i}))/(\lambda-\lambda_{i}), \nonumber\\
\eea
\noindent with the last term reducing to $(BM/2g)+\sum_{j}A_{j}\Lambda(\epsilon_{j})/(\lambda-\epsilon_{j})$. Since the eigenvalue $r_{j}$ of the conserved operator $R_{j}$ is given by the pole of $\tau(\lambda)$ at $\lambda=\epsilon_{j}$ we can read off this eigenvalue by looking at the residue of
$\tau(\lambda)$ at this pole

\bea
r_{j}=2A_{j}\Lambda(\epsilon_{j})-A_{j}(C+B)/g + 2\sum_{i\neq
j}A_{i}A_{j}/(\epsilon_{j}-\epsilon_{i}).\nonumber\\
\eea

As with any non-linear system, solving Eqs. (\ref{sequ}) necessitates an iterative method such as the well known Newton-Raphson approach. Provided we have access to an initial approximative solution that is good enough to be in its basin of attraction, convergence of the method to a specific solution will be quadratic . Finding eigenstates therefore requires a sufficiently good guess which we obtain by slightly deforming known solutions.  In the cases treated here, we exclusively know the exact eigenstates of the system at $g=0$. We therefore approach the problem by slowly deforming these $g=0$ solutions. Provided $g$ is raised in small enough steps, this guarantees that the previously found solutions can be used to generate a good approximation at the next point.

Remarkably, it is numerically simple to compute the $n$ first derivatives of the variables $\Lambda_j=g\Lambda(\e_j)$ with respect to $g$. Indeed, one can show that every order needs only the solving of the same linear system with an updated right hand side that is straightforwardly computed from the previously computed information. Defining $\Lambda_j^{(n)}=\frac{d^n\Lambda_j}{dg^n}$ we obtain in matrix form the following linear system
\bea
\mathrm{K}\vec{\Lambda}^{(n)}= \vec{R}_n
\eea
for some right hand side that depends only on the lower derivatives of $\Lambda(\epsilon_j)$ and a constant matrix
\bea
\mathrm{K}_{ij}= \begin{cases}  \frac{g}{\epsilon_i - \epsilon_j} & i\neq j \\
                  \sum_{k\neq j} \frac{-g}{\epsilon_k-\epsilon_j} + B\epsilon_j+C -2\Lambda_j & i = j
    \end{cases}.
\eea
The components of the right hand side $R_{n,j}$ can be computed iteratively as
\bea
    R_{0,j}&=&-\Lambda_j^2\nonumber\\
    R_{n,j}&=&\frac{n}{g}\left(-R_{n-1,j}+\Lambda_j^{(n-1)}\left[(B\epsilon_j+C)-2\Lambda_j\right]\right)\nonumber\\
    &&+ \sum_{k=1}^{n-1}\binom{n}{k}\Lambda_j^{(k)}\Lambda_j^{(n-k)}.
\eea
Since this requires a single matrix inversion (LU or QM decomposition to be exact) which can be reused at every order, it is a numerically a fast process to compute the $n$ first derivatives. Using the resulting derivatives, a Taylor expansion gives an excellent initial approximation to $\Lambda(\e_j)$ at $g+\Delta g$ even for fairly large $\Delta g$:

\bea
\left. \tilde{\Lambda}(\e_j)\right|_{g+\Delta g} =  \left.{\Lambda}(\e_j)\right|_{g}  + \sum_{k=1}^{n} \frac{1}{k!}  \left.\frac{\partial^k \Lambda(\e_j)}{(\partial g)^k} \right|_{g} (\Delta g)^k.
\eea

In principle, the radius of convergence of the Taylor expansion
around the current solution would set an upper limit on the $\Delta
g$ step we can take. Nonetheless, one should keep in mind that
adding terms to the Taylor series only offers linear convergence
while Newton's method converges quadratically. While computing less
derivatives and relying on more Newton steps could speed up the
calculation, it also comes with the risk that the initial guess
falls outside the basin of attraction of the desired solution.
Optimizing the computation speed without compromising stability is
then a question of balance between the number of computed
derivatives and the size of the steps one takes in $g$.

\section{Inverting $\Lambda(z)$}
\label{inverting}

Our capacity to compute physically relevant quantities relies on the use of Slavnov's determinant formula\cite{slavnov} which gives expressions for the expectation values and correlation functions in terms of simple determinants \cite{zhou} built out using the set of rapidities $\{\lambda_i\}$. It is therefore necessary to be able to extract them from a given solution obtained in terms of the variables $\{\Lambda(\e_i)\}$.

Going back to the definition of $\Lambda(z)$ (Eq. \ref{sdef}), Newton's identities allow us to write it explicitly in terms of elementary symmetric polynomials. We have

\bea
\Lambda(\e_j) = \frac{P'(\e_j)}{P(\e_j)} = \frac{\displaystyle\sum_{m=0}^M m \e_j^{m-1} P_{M-m}}{\displaystyle \sum_{m=0}^M \e_j^m P_{M-m}},
\eea

\noindent with

\bea
P_k &=& (-1)^k  \sum_{1\le j_1< j_2 < \dots < j_k \le M}\lambda_{j_1}  \lambda_{j_2} \dots \lambda_{j_k}
\nonumber\\
P_0 &=& 1.
\eea

Having the $N$ values $\Lambda(\e_j)$ at hand, writing a linear problem for the elementary symmetric polynomials is then a trivial matter:

\bea
\displaystyle\sum_{m=0}^{M-1} \left(m \e_j^{m-1} - \Lambda(\e_j) \e_j^m \right) P_{M-m} &=&
\Lambda(\e_j) \e_j^M -M \e_j^{M-1}. \nonumber\\
\eea

\subsection{Case $M < N$}
\label{mltn}

When the number of excitations $M$ is smaller than $N$, one simply needs to pick $M$ values of $(\e_j)$ to extract the polynomials $P_{M-m}$. The complete set of $M$ elementary symmetric polynomials give the real coefficients of the single variable polynomial $P(z)$ at every order. The corresponding ensemble of parameters $\{\lambda_i\}$ which defines the eigenstate is then obtained by finding every roots of $P(z)$.

This is a well studied problem for which many methods exists. Here we choose to use Laguerre's method with polynomial deflation. Laguerre's method is an almost "sure-fire" method for finding one root $r_1$. We then proceed by finding a root $r_2$ of the deflated polynomial $\frac{P(z)}{z-r_1}$ at the next step, then a root of $\frac{P(z)}{(z-r_1)(z-r_2)}$, etc. Repeating $M$ times allows the extraction of every root of our initial polynomial, i.e. every rapidity $\lambda_i$.

This procedure for inverting the $\Lambda(z)$ function is purely local in $g$ in the sense that it only needs $\Lambda(\epsilon_j)$ at a given $g$. Therefore it is not necessary to perform the inversion at values of $g$ we are not interested in. Solving for $\Lambda(\epsilon_j)$ still requires this scan in $g$ but if we are, for example, only interested in the large $g$ regime, not having to perform the inversion at intermediate steps can markedly reduce computation time.

\subsection{Case $M > N$}
\label{mgtn}

For a Hamiltonian realized in terms of only spin 1/2 degrees of freedom the number of excitations is always $M \le N$. However, in more general cases there exists a subspace where the excitation number is larger than the number of degrees of freedom ($M>N$). In such a case, one cannot simply use the set of $N$ parameters $\Lambda(\e_j)$ to find the $M > N$ elementary symmetric polynomials.

However, for any solutions with $M \le \sum_{i=1}^N 2 A_i$, one can still extract them by solving only linear problems, since the supplementary information has been obtained from the necessity to solve for certain derivatives $\Lambda'(\e_i)$. The unbounded number of excitations in models containing bosonic degrees of freedom such as the Dicke models presented here, would give a subset of states with $M > \sum_{i=1}^N 2 A_i$. These would not allow such a simple inversion. However, since any degree of freedom with $A_i$ can only accommodate up to $ 2 A_i$ excitations, for any model based exclusively on spins (of any magnitude and degeneracy) the total number of excitations is always bounded and every solution has $M \le \sum_{i=1}^N 2 A_i$. 

In solving the quadratic equations (\ref{eq1},\ref{eq2},...) we are provided with $\Lambda(\e_i)$ as well as its $2 A_i-1$ first derivatives. Having these values at hand, writing a linear problem for the elementary symmetric polynomials is simply a matter of defining rational functions which are linear in $P_n$ in both the numerator and denominator. Naturally, we already have from the definition

\bea
\Lambda(z) = \frac{P'(z)}{P(z)} = \frac{\displaystyle\sum_{m=0}^N m z^{m-1} P_{N-m}}{\displaystyle \sum_{m=0}^N z^m P_{N-m}},
\eea

but we can also write

\bea
V_1(z) \equiv \Lambda'(z) + \Lambda(z)^2 &=&   \frac{P''(z)}{P(z)} ,
\nonumber\\
V_2(z) \equiv \Lambda''(z)+3\Lambda(z) \Lambda'(z) +\Lambda(z)^3 &=&  \frac{P'''(z)}{P(z)} ,
\nonumber\\
\dots
\eea

By evaluating the $M$ necessary functions $(\Lambda(z),V_1(z),V_2(z), ...) $ at $z=\e_j$, we can write $M$ equations linear in the polynomials $P_n$. Solving them and using a root finding algorithm for $P(z)$ would, once again, give us the set of rapidities $\{\lambda_i\}$.

\section{Applications}
\label{applications}

We now turn to three specific models derived from a generalized Gaudin algebra in order to demonstrate the efficiency of this approach. First we treat the discrete reduced BCS\cite{bcs-57} model (Richardson model). One of the authors has already been involved in solving the Bethe equations in this context \cite{fcc-08,fcc-10,fcc-09,fcc-09p}. However, in this series of papers, it was performed without the currently discussed method. This led to serious stability and computation time issues which are now extremely well controlled in the current approach.

\subsection{Richardson model}
\label{richardson}

The Richardson model\cite{rs-62}, while having a rich history in nuclear physics, has also recently found renewed interest in condensed matter physics in light of tunneling experiments performed on superconducting nanograins \cite{exp}. It is nothing but a discrete version of the reduced BCS model which limits the interaction to a uniform s-wave pairing term between time reversed states. Using Anderson pseudospin-$\frac{1}{2}$ representation in terms of fermionic operators

\bea
S_i^z &=& c_{i\uparrow}^\dag c_{i\downarrow}^\dag c_{i\downarrow} c_{i\uparrow} -\frac{1}{2}
\nonumber\\
S_i^+ &=& c_{i\uparrow}^\dag c_{i\downarrow}^\dag
\nonumber\\
S_i^- &=& c_{i\downarrow} c_{i\uparrow},
\eea

\noindent the Hamiltonian is given by

\bea
H= \sum_{i=1}^N\e_i S^z_i - g \sum_{i,j = 1}^NS^+_i S^-_j.
\label{hbcs}
\eea

Here $\epsilon_j$  corresponds to the discrete set of unblocked single fermion energies which can accommodate a Cooper pair, while $g$ is the pairing strength between time-reversed fermionic states. While its integrability in a strict sense was proven much later\cite{crs-97}, the exact solution of the model was first proposed by Richardson himself \cite{rs-62}. The eigenstates of the system are of the form given in Eq. (\ref{bethestate}) with a pseudovacuum $\left|0\right>= \left|\downarrow,\downarrow,..., \downarrow\right>$ that has fully down polarized pseudospins (Fock vacuum of Cooper pairs ) and the Gaudin creation operators given by

\bea
\mathrm{S}^+(\lambda_\a) = \sum_{i=1}^N \frac{S^+_i}{\lambda_\a - \e_i}.
\label{s+bcs}
\eea

The Bethe equations for this particular realization are given by\cite{aff-01,dp-02}

\be
\sum_{\a=1}^N\frac{1/2}{\lambda_j - \e_\a}+\frac1{2g}=\sum_{k\neq j}^{M}
\frac1{\lambda_j-\lambda_k}\,\quad
j=1,\dots, M\,.
\label{RICHEQ}
\ee

This obviously corresponds to an $F$ function of the type given in Eq. (\ref{eq:ffunction}) with $A_i = \frac{1}{2}$, $B=0$ and $C = 1$. Solving the model is straightforwardly carried out using the procedure outlined in this article. For any set of rapidities which is a solution of the Bethe equations, the eigenenergy of the state is then given by

\bea
E = \sum_{j=1}^N \lambda_j.
\eea

At zero-coupling, the $M$-pairs eigenstates are simply obtained occupying any $M$ energy levels  with a Cooper pair (pseudospin up) while the remaining $N-M$ are empty. This can be represented by the set of rapidities $\lambda_\alpha = \epsilon_{i_\alpha}$ which according to Eq. (\ref{s+bcs}) lead to an up pseudospin at every energy $\epsilon_{i_\alpha}$ present in the set of rapidities. This state obviously leads to diverging $\displaystyle \Lambda(\e_{i_\a}) = \sum_{\a} \frac{1}{\e_{i_\a}-\lambda_\alpha}$ for the occupied levels, but linearizing the Bethe Eqs. (\ref{RICHEQ}) tells us that every rapidity can be written as $\lambda_i = \e_{\a_i} - g$ at weak coupling. We consequently solve Eqs. (\ref{sequ}) using $\Lambda_j = g\Lambda(\e_j)$ which are given at zero coupling by

\bea
\Lambda_j =
\left\{
\begin{array}{cc}
1& \ \ \ \mathrm{occupied} \ \ \e_j     \\
0 & \ \ \ \mathrm{empty} \ \ \e_j.
  \end{array}
\right.
\eea

\begin{figure}[h]
\includegraphics[width=4.2cm]{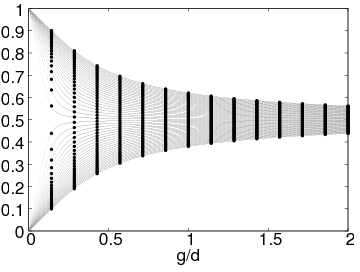}\includegraphics[width=4.2cm]{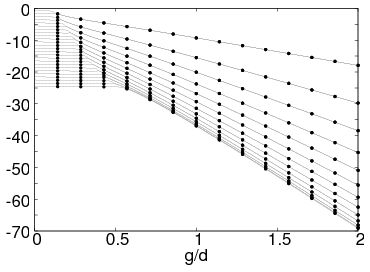}
\includegraphics[width=4.2cm]{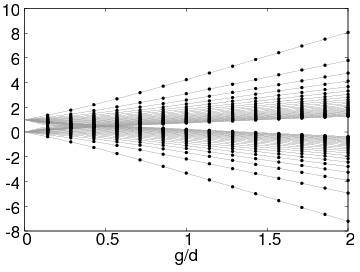}\includegraphics[width=4.2cm]{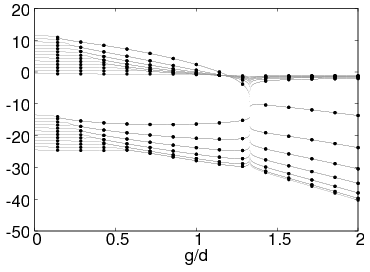}
\includegraphics[width=4.2cm]{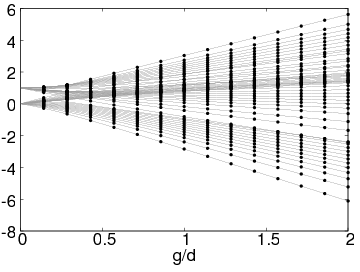}\includegraphics[width=4.2cm]{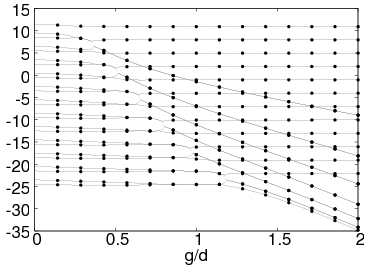}
\caption{$g \Lambda(\e_j)$ (left) and the real part of the corresponding rapidities (right) for the ground state (top) and two excited states of the equally spaced Richardson model (N=50, M=25). The circles are the only points needed when using the 6 first derivatives for the regression.}
\label{raps}
\end{figure}

While in other applications, such as nuclear physics \cite{dh-03,zv-05}, a different distribution of the levels could be better suited, we present results for equally spaced single particle energy levels within a bandwidth $D$, i.e. $\epsilon_j = - \frac{D}{2}+ \frac{D}{N} \ j$. This Richardson model is the one typically used in condensed matter systems\cite{dr-01}. Both $\Lambda_j = g\Lambda(\e_j)$ and the corresponding rapidities for the ground state and two excited states are plotted in Fig. \ref{raps}.

The black circles show the points at which we actually solve the equations, the step size in $g$ between two points being $\Delta_g/d = \frac{1}{7}$. These results were computed using the 6 first derivatives for the regression (see section \ref{solving}). For reference we present in gray the numerical solution as a continuous function of $g/d$. The figure therefore shows how this approach allows steps in the coupling constant which are very large compared to the actual scale on which rapidities themselves vary.

The middle panel shows a state specifically chosen to generate a region of rapid variation of the rapidities that one can see around $g/d = 1.3$. When solving directly for the rapidities themselves using Eqs. (\ref{RICHEQ}), one would need an extremely small step size in this region in order to maintain stability. In the current approach however, this complex structure causes no problem since the variables $\Lambda_j$ remain smoothly varying functions. This constitutes a prime example of the remarkable capacities of this new approach which allows the scan in $g$ to be performed orders of magnitude faster than using the standard Bethe Eqs. (\ref{RICHEQ}).

\subsection{Generalized Dicke model}
\label{dicke}

We now turn to a second related model, the Generalized Dicke model. It describes a collection of $N$ multi-level systems coupled uniformly to a single bosonic mode and has been shown to be  Bethe Ansatz solvable\cite{Gaudin_paper,dicke_derive}. Here, we again consider the case of 2-level systems (represented by a spin 1/2) which is evidently the relevant  case for cavity-based quantum computing proposals\cite{qed}. The Hamiltonian takes the form
\bea
H=\omega b^{\dag}b + \sum_{j=1}^N \epsilon_j S_j^z + V \sum_{j=1}^N \left( b^{\dag} S_j^- + S_j^+b\right),
\eea
where $\{S_j^z,S_j^+,S_j^-\}$ are the usual generators of spin 1/2 representations of SU(2) at each site $j$.   The boson frequency is $\omega$, $\e_j$ sets the splitting between both levels of every subsystem, while $V$ controls the strength of the interaction. This Hamiltonian conserves the total number of excitations $ b^{\dag}b + \sum_j\left(S_j^z +\frac{1}{2}\right)$. The construction of $M$-excitations eigenstates is achieved using the Gaudin creation operators
\bea
\mathrm{S}^+(\lambda_\a)=b^{\dag} + \sum_j \frac{V}{\lambda_{\alpha} - \epsilon_j} S_j^+
\label{DICKEEIGENSTATES}
\eea
acting repeatedly on the pseudo-vacuum state $\left| 0 \right> = \left|0;\downarrow,\downarrow,..., \downarrow\right>$ which contains no boson and is fully down-polarized. The appropriate sets of rapidities  $\{\lambda_{\alpha}\}_{\alpha=1}^M$ must fulfill the Bethe equations \cite{jan-10}
\bea
\frac{\omega}{2V^2}-\frac{\lambda_{\alpha}}{2V^2}-\sum_{i=1}^{N}\frac{1/2}{\epsilon_i-\lambda_{\alpha}}=\sum_{\beta\neq\alpha}^{M}
\frac{1}{\lambda_{\alpha}-\lambda_{\beta}}.
\label{DICKEEQ}
\eea
Here, we can identify the general form (Eq. \ref{eq:ffunction}) with $g=V^2$, $A_i\equiv 1/2$, $B=-1$ and $C=\omega$. The energy eigenvalues corresponding to each set of rapidities are
\bea
E(\{\lambda_\alpha\}) = \sum_{\alpha} \lambda_{\alpha}-\sum_j \epsilon_j.
\eea
Now we turn to the solution of (\ref{DICKEEQ}) using the method proposed in the previous section. The equivalent equations for $\Lambda_j=V^2\Lambda(\epsilon_j)$ read
\bea
    V^2\sum_{i\neq j} \frac{\Lambda_i-\Lambda_j}{\epsilon_i-\epsilon_j}  =  \Lambda_j \left[\left(\epsilon_j-\omega\right)  + \Lambda_j\right] - V^2M.
\label{DICKESUBS}
\eea

By linearizing the Bethe equations (\ref{DICKEEQ}), one can show that $\lambda_{\alpha} =\epsilon_j - \frac{V^2}{\omega-\epsilon_j}$ give correct solutions in the limit $V^2\rightarrow 0$. These values lead to $\Lambda_j = \omega-\epsilon_j$ at $V=0$. As in the Richardson model, $\lambda_{\alpha} =\epsilon_{j_\a}$ leads to an up-spin at level $\epsilon_{j_\a}$.
Another possible solution is obtained for $\lambda_\alpha = \omega$ which leads to $\Lambda_j = 0$. In this case, every $\lambda_{\alpha}=\omega$ corresponds to an additional  bosonic excitation at $V^2=0$, since eigenstates are constructed using the operator (\ref{DICKEEIGENSTATES}). Any of the $d=\sum_{i=1}^M \binom{N}{i}$ combinations of $M$ excited spins and bosons gives us a possible eigenstate of $H$ at $V^2=0$, which we deform numerically to find eigenstates and eigenvalues at nonzero coupling $V^2>0$.

\begin{figure}[h]
\includegraphics[width=4.2cm]{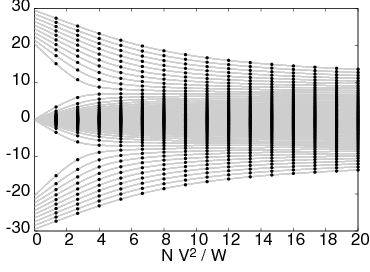}\includegraphics[width=4.2cm]{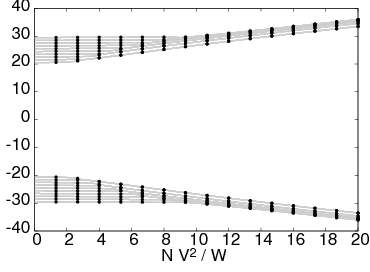}
\includegraphics[width=4.2cm]{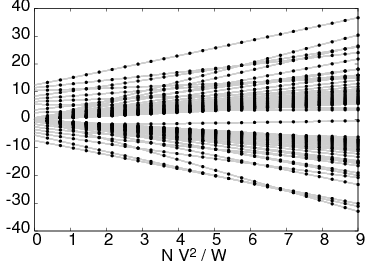}\includegraphics[width=4.2cm]{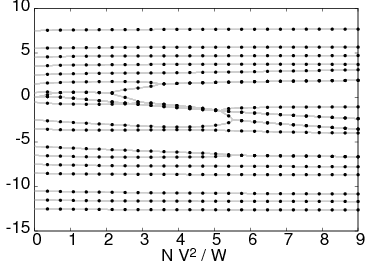}
\caption{$V^2 \Lambda(\e_j)$ (left) and the real part of the corresponding rapidities (right) for the "equally spaced" Generalized Dicke model (N=60, M=20). The circles are the only points needed when using the 5 first derivatives for the regression.}
\label{raps_dicke}
\end{figure}

Figure \ref{raps_dicke} shows the rapidities and $\Lambda_j$ for two different states as a function of $V^2$. We made a specific choice for the inhomogeneity by using equally spaced splittings $\epsilon_j = -\frac{D}{2}+ \frac{D}{N} \ j$. This would ultimately correspond to a coarse-graining of a flat distribution of splittings within a bandwidth $D$. Moreover, we choose the bosonic frequency to be at the midpoint of the band, i.e. $\omega = 0$.

\subsection{Central Spin model}
\label{centralspin}

Finally, we apply the method to the Central Spin (CS) model which in its most frequently encountered form describes a single spin coupled to an external magnetic field and via hyperfine interaction, to a spin bath. It is straightforwardly obtained from the commuting conserved quantities of the SU(2) XXX Gaudin models given by

\bea
R_i = S^z_i + \sum_{j \ne i} \frac{g}{\epsilon_i-\epsilon_j} \vec{S}_i \cdot \vec{S}_j,
\eea
\noindent by defining the exactly solvable model as $H=\frac{R_0}{g}$, with the choice $\e_0 = 0$, $h_0 = \frac{1}{g}$ and $A_j=-\frac{1}{\epsilon_j}$ , i.e.

\bea
H = h_0 S^z_0 + \sum_{j \ne 0} A_j \vec{S}_0 \cdot \vec{S}_j.
\label{eq:csham}
\eea

While the external magnetic field is only coupled to the central spin $i=0$, the conservation of the total excitation number $S^z_0 +\displaystyle \sum_{j\ne 0} S^z_j$ makes the model equivalent to

\bea
H = h g_e S^z_0 + h g_N \sum_{j \ne 0}  S^z_j  + \sum_{j \ne 0} A_j \vec{S}_0 \cdot \vec{S}_j,
\eea

\noindent which would also include coupling of the bath spins to the external magnetic field.

 In principle, any linear combination of the integrals of motion also leads to an integrable model which commutes with Hamiltonian (\ref{eq:csham}) and therefore shares the same eigenbasis. Consequently any Hamiltonian

 \bea
 H &=& \sum_{i} \eta_i R_i = \sum_i h \eta_i S^z_i + \sum_{i, j > i} \frac{\eta_i - \eta_j}{\epsilon_i-\epsilon_j} \vec{S}_i \cdot \vec{S}_j \nonumber\\ &\equiv& h_0 S^z_0 + \sum_{i\ne0} h_i S^z_i + \sum_{i\ne0} A_{i} \vec{S}_0 \cdot \vec{S}_i+ \sum_{i\ne0, j > i} B_{i,j} \vec{S}_i \cdot \vec{S}_j,\nonumber\\
 \eea

 \noindent is also Bethe Ansatz solvable. While this allows the treatment of models which would include hyperfine interactions between the bath spins $i > 0$, integrability restricts the allowed values of the couplings. For example, if we suppose that bath spins are homogeneously coupled to the external field, i.e. $\eta_j = \eta \ \ \forall j > 0$, integrability imposes the absence of coupling between bath spins. On the other hand, models with non-zero interaction in the bath can be built from any choice of $\eta_j$ but this in turn would force the presence of an inhomogeneous magnetic field acting differently on every bath spin.

 In the limit $h = 0$, one could study models with bath interactions provided the couplings respect the constraints

 \bea
\frac{B_{jk}}{\eta_j-\eta_k}(A_j(\eta_0-\eta_k)-A_k(\eta_0-\eta_j))=A_jA_k,
\label{constr}
\eea

\noindent which has important consequences. For example when any two nuclear spins are equally coupled ($A_j=A_k=A$) to the central one, the coupling between those two spins  would also need to be given by $B_{jk} = A$. For a given set of distinct $A_j$, it would however still be possible and interesting to study Central Spin systems with uniform bath couplings $B_{i,j}=B$ or any integrable case that emerges from a given choice of distinct $\eta_j$.

We still choose to focus on Hamiltonians of the form (\ref{eq:csham}). Is is straightforward to show that it commutes with the Richardson Hamiltonian (Eq. \ref{hbcs}) defined by parameters $g = -\frac{1}{h}, \epsilon_0 = 0,\epsilon_j = -\frac{1}{A_j}$. They therefore share common eigenstates so that, in principle, solving Eqs. (\ref{RICHEQ}) for negative values of $g$ would give us the eigenstates of the central spin Hamiltonian. However, we follow the alternative road of inverting the spin quantization axis ($\hat{z} \to -\hat{z}$) and solving Eqs. (\ref{RICHEQ}) for positive $g$. Eigenstates are then be obtained using the operators

\bea
S^+(\lambda)=\sum_{j=0}^{N_b}\frac{S^-_j}{\lambda-\epsilon_j},
\eea

\noindent acting on the fully up-polarized pseudo-vacuum $\left|0\right> = \ket{\uparrow\uparrow\dots\uparrow}$ and the eigenenergy of a given eigenstate is given by
\bea
E_0(\{\lambda_k\}_{k=1,\dots,M})=\sum_{j=1}^{N_b}\frac{1}{\epsilon_0-\epsilon_j}+2\sum_{k=1}^M\frac{1}{\lambda_k-\epsilon_0}+\frac{h}{2}.\nonumber\\
\eea

In some of the physically relevant systems that can be described with Central Spin model (such as Quantum Dots or Nitrogen-Vacancies (NV) in diamond\cite{NVa}), the proper interactions between the CS and the nuclei do not give rise to equally spaced values of $\epsilon_j = -\frac{1}{A_j}$. In fact, some of the parameters $\epsilon_j$ can be very close to one another while others are strongly separated. While the Bethe equations are exactly the same as for the Richardson model this distribution of $\e_j$ has a direct consequence on the application of the current method. Since for equally spaced levels with separation $d$, the Bethe equations can be written only in terms of $g/d$, the size of the steps in $g$ that one can use is simply controlled by the $d$ parameter. However, when a number of levels are very close by while others are far from one another, these variable spacings lead to variations of $\Lambda(\e_j)$ controlled by different scales. Naturally, using a step in $g$ that is a given fraction of the smallest distance $\e_j-\e_{j+1}$ would insure stability but it would ultimately necessitate a large number of points to reach the strong $g=\frac{1}{h}$ limit (weak magnetic field). It is therefore beneficial to use a variable step size. Starting from small steps at low $g$, we turn to larger ones when $g$ has increased sufficiently and the behavior of the solutions vary on a much slower scale. Variables steps could also be used in the previously studied models, but in the current case it is more or less necessary to do so in order to achieve fast computation.

\begin{figure}[h]
\includegraphics[width=4.2cm]{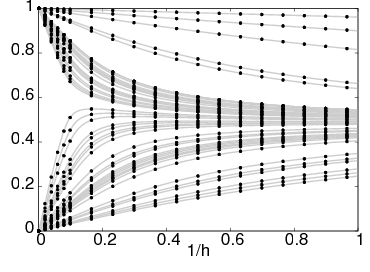}  \includegraphics[width=4.2cm]{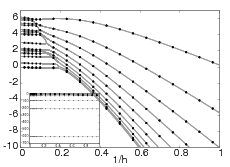}%
\\
   \includegraphics[width=4.2cm]{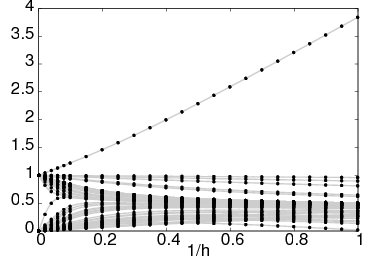}  \includegraphics[width=4.2cm]{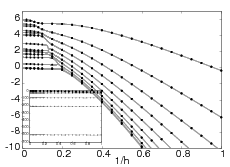}%
\caption{
$g\Lambda(\epsilon_j)$ (left) and the rapidities (right) for the dipole couplings of an NV-Center \cite{lukin} $\epsilon_j=-4/A_j$ and $A_j=5.6(\frac{a_{jj}}{R_j})^3(1-3\cos{(\theta_j)}^2)$[MHz], where $a_{jj}=1.54$[A] is the nearest neighbor distance for diamond and $R_j$ is the distance between the j-th carbon atom and the defect (center).
$\theta_j$ is the angle between the magnetic field and $\vec{R}_j$. The vectors $\vec{R}_j$ are  corrected by a small amount of randomness to avoid degeneracies.
Top panel shows the ground state; bottom one shows an excited state. Insets show the full range of rapidities while the main figures focus on the region with non-trivial structure in the given range of magnetic field.}
\label{rapshom}
\end{figure}

Figure (\ref{rapshom}) shows the behavior of the rapidities for an NV-Center ($N=50$ and $M=25$), in the interval $g$ going from $0$ to $0.1$ the Newton steps were taken for $gS(\epsilon_j)$ with a step size $\Delta g=1/50$, for $g>0.1$ the step size is increased to $\Delta g=1/15$ ($\Delta g=1/20$ for the excited state). The first $6$ derivatives are used.

Due to the structure of the levels $\epsilon_j$ we choose to restrict the y-axis in the plots of the rapidities to make the intricate structure of the solutions visible. The insets present them in the complete range.

Once again, the black dots represent the points at which the equations are solved. One can notice the increase in the step size at large $g \equiv \frac{1}{h}$. In spite of the necessity of using smaller steps, these plots show without a doubt that this approach still allows us to realize the scan in $g$ using large steps on the scale on which rapidities themselves vary.

A variable step-size could naturally be defined by making use of the fact that we compute the truncated Taylor expansion up to a given order. One could fix the appropriate steps simply by defining a desired level of precision for the limited expansion. This would allow one to control stability,  while adapting the steps to the behavior of the $\Lambda(\e_j)$ variables at the current $g$ point.

\section{Conclusions}

In this work we have shown how, for rational Gaudin models, the set of non-linear coupled Bethe equations can be transformed to a new set of equations which is significantly simpler to solve numerically. We presented the complete algorithm and applied it to a variety of models derived from the generalized Gaudin algebra.

This work exclusively presents and demonstrates the capacities of this new approach to the solving of Bethe equations. In light of previous work \cite{fcc-08,fcc-10,fcc-09,fcc-09p}, it should however be clear that by allowing rapid and systematic solving of a decent fraction of the full Hilbert this can directly be used for relevant physical calculations. For example, studies of the non-equilibrium dynamics of the discussed models are currently pursued by the authors. The outcome of these calculations should prove invaluable considering that this technique gives direct access to regimes which have been hard to describe before. To name only this one, the weak magnetic field limit of the Central spin models is a prime example.

While we focused on rational Gaudin models, algorithms tailored to the trigonometric or hyperbolic XXZ\cite{Ortiz} models or general XYZ\cite{XYZ} could possibly be built along similar lines. This is a more fundamental question that is left open for the time being.

\section*{Acknowledgments}

AF and CS's work was supported by the DFG through SFB631, SFB-TR12
and the Excellence Cluster "Nanosystems Initiative Munich (NIM)".
OEA and VG acknowledge discussions with P. Barmettler and support of
the the Swiss National Science Foundation.

\end{document}